\documentclass[useAMS,usenatbib]{mn2e}

\usepackage{graphicx}   
\usepackage{amsmath}    
\usepackage{amssymb}    
\usepackage{threeparttable}
\usepackage{supertabular}
\usepackage{longtable}
\usepackage{booktabs}

\title[Near Mean Motion Resonance of Terrestrial Planets]{Near Mean Motion Resonance of Terrestrial Planet Pair Induced by Giant Planet: Application to Kepler-68 System}

\author[Mengrui Pan et al.]{
Mengrui Pan,$^{1,2}$  Su Wang,$^{1}$\thanks{} Jianghui
Ji$^{1,2,3}$\thanks{E-mail: jijh@pmo.ac.cn, wangsu@pmo.ac.cn}
\\
$^{1}$CAS Key Laboratory of Planetary Sciences, Purple Mountain Observatory, Chinese Academy of Sciences, Nanjing 210008, China\\
$^{2}$School of Astronomy and Space Science, University of Science and Technology of China, Hefei 230026, China\\
$^{3}$CAS Center for Excellence in Comparative Planetology, Hefei 230026, China\\
}

\pubyear{2020}

\begin{document}
\label{firstpage}
\pagerange{\pageref{firstpage}--\pageref{lastpage}} \maketitle

\begin{abstract}
In this work, we investigate configuration formation of two inner
terrestrial planets near mean motion resonance (MMRs) induced by the
perturbation of a distant gas-giant for the Kepler-68 system, by
conducting thousands of numerical simulations. The results show that
the formation of terrestrial planets is relevant to the speed of
Type I migration, the mass of planets, and the existence of giant
planet. The mass and eccentricity of the giant planet may play a
crucial role in shaping the final configuration of the system. The
inner planet pair can be trapped in 5:3 or 7:4 MMRs if the giant
planet revolves the central star with an eccentric orbit, which is
similar to the observed configuration of Kepler-68. Moreover, we
find that the eccentricity of the middle planet can be excited to
roughly 0.2 if the giant planet is more massive than 5 $M_J$,
otherwise the terrestrial planets are inclined to remain
near-circular orbits. Our study may provide a likely formation
scenario for the planetary systems that harbor several terrestrial
planets near MMRs inside and one gas-giant exterior to them.
\end{abstract}


\section{Introduction} \label{sec:intro}
\begin{figure}
\includegraphics[width=\columnwidth,height=6cm]{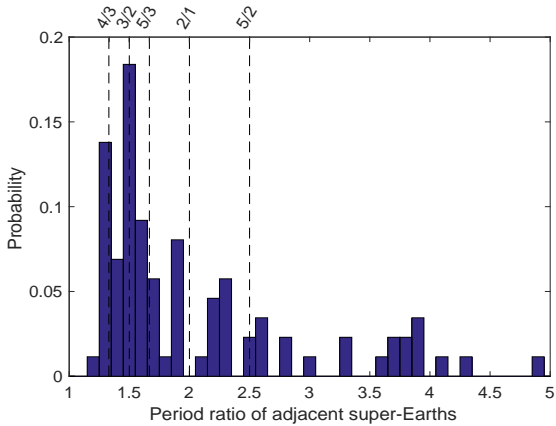}
\caption{Distribution of period ratio of two adjacent super-Earths
in exoplanetary systems. \label{fig:pratio}}
\end{figure}

Kepler space telescope discovered a great number of tightly packed
terrestrial planet pairs, which are involved in or near mean motion
resonances (MMRs) \citep{Lissauer2011a, Ford2012, Rowe2014,
Gozdziewski2016, Berger2018}. From a viewpoint of statistics, for
Kepler planetary candidates, there are two peaks at the distribution
of period ratio of two adjacent terrestrial planets near 1.5 and
2.0, respectively \citep{Lissauer2011b, Fabrycky2014, Gillon2017,
Charal2018}. Figure \ref{fig:pratio} shows the distribution of
period ratios of adjacent terrestrial planet pairs whose masses are
confirmed (Herein we refer to the terrestrial planet with a mass
$M_p < 10~M_\oplus$), where 87 terrestrial planet pairs are
included. Most of them locate close to central star. The entire
distribution of period ratios illustrates that the planet pairs have
a pileup around 5:3, 3:2, 2:1, 5:2 and 3:1 MMRs. Combined with
abundant observations, the near-resonant terrestrial planet pairs,
accompanied by one or more gas-giants, are also discovered in a
couple of planetary systems, e.g., Kepler-48, Kepler-68 and
Kepler-154 \citep{Marcy2014, Uehara2016, Mills2019}. Thus, this
leads to the crucial questions: what scenario may produce the
configuration of terrestrial planet pairs near MMRs, and how the
formation of such configuration can be affected if there exists an
additional gas-giant exterior to the terrestrial planet pair?

Several scenarios, such as in situ formation \citep{Chiang2013},
inside-out formation \citep{Chatterjee2013}, pebble-accretion
\citep{Liu2018,Liu2019}, or late orbital instability after disk
depletion \citep{Izi2017, Ogihara2018, Lam2019} have been suggested
to explain the formation of hot super-Earth in or near MMR
configuration. However, \citet{Ogihara2015} claimed close-in
super-Earths cannot be formed in situ, unless their migration speed
is suppressed in the entire disk inside 1 AU. The widely-accepted
theory indicates that the short-period planets are formed at large
distance far away from their host stars and migrate to currently
observed orbits through angular momentum exchange with
protoplanetary disk \citep{Lin1986, Ward1997, Raymond2018}. A large
majority of planets in compacted systems are super-Earths that can
be easily trapped into MMRs through Type I migration
\citep{Cresswell2006}. In addition, previous studies proposed an
alternative scenario to shed light on the formation of near-resonant
systems via Type I migration \citep{Wang2012, Wang2014, Wang2017}.
As a matter of fact, most of the confirmed planet pairs are not
trapped in exact MMRs. The stellar magnetic field, inviscid disks or
turbulence in protoplanetary disk triggered by magneto-rotational
instability stochastic can reproduce the distribution of period
ratio bearing resemblance to nowadays observations of the planet
pair \citep{Rein2012, Paardek2013, Liu2017, McNally2019, Liu2019}.
Other mechanism suggests that planet-planet scattering and final
mergers of several core/planets in the cavity tend to disrupt the
MMRs established during the migration \citep{Terquem2007}.
Furthermore, for those systems that tidal dissipation is
unexpectedly efficient, weak dissipation damps the eccentricities of
planets, thereby driving the near-resonant pairs move out of
resonance \citep{Lithwick2012, Lee2013}. In summary, the above
scenarios give some explanations on the formation of system with
terrestrial planets.

\begin{table*}
\caption{Orbital Parameters of the Kepler-68 Planetary System.}
\centering \label{tab:k68}
\begin{tabular}{cccccc}
\hline
Planet& Mass & Semi-major axis & \multicolumn{2}{c}{Eccentricity} & Orbital Period \\
\cline{4-5}
 & ($\rm M_{\oplus}$) & ($\rm AU$) & (1) &(2) &$ (\rm day$)\\

   \hline
   b & $5.98^{+1.70}_{-1.70}$ & $0.0617^{+0.00056}_{-0.00056}$ & 0.0 & $0.02^{+0.13}_{-0.02}$ & $5.399^{+0.000004}_{-0.000004}$ \\
   \specialrule{0em}{3.0pt}{3.0pt}
   c & $2.18^{+0.011}_{-0.011}$ & $0.09059^{+0.00082}_{-0.00082}$ & 0.0 & $0.42^{+0.41}_{-0.1}$ & $9.605^{+0.000072}_{-0.000072}$ \\
    \specialrule{0em}{3.0pt}{3.0pt}
   d & $308.5^{+11.124}_{-11.124}$ & $1.4^{+0.03}_{-0.03}$ & $0.18^{+0.05}_{-0.05}$ & - & $580^{+15}_{-15}$ \\
   \hline
\end{tabular}
\begin{tablenotes}
\footnotesize
\item[]{Note: The parameters are adopted from  https://exoplanetarchive.ipac.caltech.edu/index.html, \citet{Gilliland2013},  \citet{Mills2019}, \citet{VanEylen2015}. }
\end{tablenotes}
\end{table*}

So far, a large number of systems, which host giant planets involved
in MMRs, have been extensively investigated \citep{Lee2002,
Robertson2012, Lee2013, Marti2013, Wittenmyer2016, Marti2016,
Bae2019}.  Figure \ref{fig:pratio} contains nine systems with the
co-existence of terrestrial planets and gas-giant (with a mass above
0.1 $M_J$).  The planetary embryos can be accelerated to collide and
merge into terrestrial cores, further to produce super-Earth size,
close-in terrestrial planets under the circumstance of an additional
gas-giant in the system \citep{Zhou2005, Raymond2006, Mandell2007,
Hands2016, Sun2017}. For instance, \citet{Hands2016} suggested the
outer planet, which can undergo exponential growth up into a giant
planet, tends to push interior super-Earths into more tightly
first-order resonant orbits. \citet{Sun2017} revealed that MMRs
configuration can occur between gas-giants and terrestrial planets.
\citet{Granados2018} showed that an undetected outer giant planet
may have an effect on the stability and resultant configuration of
tightly packed inner planets by secular resonance. Therefore, we
believe that the existence of giant planet, in fact, does play a
significant role in sculpting final configuration of the system.

The objective of the present work is to study the influence of a
faraway gas-giant on the configuration formation of terrestrial
planets especially for the planet pair in or near MMRs. Here we take
Kepler-68 system as a template, which consists of two inner
terrestrial planets involved in near MMRs and a distant giant
companion \citep{Gilliland2013, Stassun2017}. Kepler-68 is a
solar-mass ($\sim{1.08}~M_{\sun}$) star with effective temperature
$\sim$ 5793 K \citep{Gilliland2013, Stassun2017}. Two transiting
Earth-sized planets were discovered around the star at a distance of
0.0617 and 0.09059 AU, respectively. The period ratio of the inner
planet pair is roughly 1.779, which is between 1.5 and 2.0. An
additional Jupiter-mass planet, which locates exterior to two
terrestrial planets at 1.4 AU, was detected by radial velocity
\citep{Gilliland2013, Marcy2014} (see Table 1). Consequently, this
enables us to improve the understanding of formation of
near-resonant terrestrial planets.

One of the most noteworthy is the eccentricities of Kepler-68 b and
c. They were constrained to be zero by modelling
\citep{Gilliland2013, Mills2019}, while \citet{VanEylen2015} pointed
out that the eccentricity of Kepler-68 c could amount up to 0.42
based on high-quality Kepler transit observations, which can remain
steady for $10^8$ years in the simulation.

The formation scenario of late orbital instability may be one of
possible mechanism to explain the formation of the inner
configuration of two terrestrial planets \citep{Izi2017,
Ogihara2018, Lam2019}. After the late orbital instability, the
typical orbital separation between planets is about $20~R_H$, which
is comparable to the current orbital separation between $Pb$ and $Pc$ in
the Kepler-68 system. In addition, as the period ratio of $Pb$ and $Pc$
is 1.78, which is out of the exact 5:3 resonance by more than 5\%,
this system can be regarded as a non-resonant system. The late
orbital instability can explain the formation of non-resonant
systems. Furthermore, the eccentricity of $Pc$ can be excited to some
extent during the late orbital instability. In consideration of the
existence of giant planet in the outer region, here we aim to figure
out the configuration formation of the inner planet pair, and
further investigate the excitation of eccentricity of the middle
planet under the influence of a giant planet lying outside in
Kepler-68 system. In this work, the inner planet pair can be trapped
in 5:3 or 7:4 MMRs if the giant planet moves around the central star
in an eccentric orbit, which is similar to the observed orbits,
deviating from the exact 2:1 or 3:2 MMRs. Moreover, we find that if
the giant planet is more massive than 5 $M_J$, the eccentricity of
the middle planet can be excited to approximately 0.2.

This paper is organized as follows. In Section \ref{sec:models}, we
present the gas disk model and Type I migration scenario adopted in
our simulations. In Section \ref{sec:results}, we show the numerical
simulation results with respect to the Kepler-68 system. We
summarize major conclusions in Section \ref{sec:summary}.

\section{Models} \label{sec:models}
\subsection{Disk models} \label{subsec:disk model}
To explore the configuration formation of Kepler-68 system, we make
an assumption that planets are formed in the protoplanetary disks.
Here we take empirical Minimum-Mass Solar Nebular (MMSN) model
\citep{Hayashi1981} as the gas disk model. Thus the density at a
stellar distance $a$ can be described as follows
\begin{equation} \label{equ:sigma}
\Sigma_{\rm g}=\Sigma_0(\frac{a}{1 \rm
AU})^{-k}\exp(-\frac{t}{t_n}),
\end{equation}
where $\Sigma_0=1700 \rm \ g \ cm^{-2}$ is the initial density of
the gas disk. The disk model can be shallower and the surface
density can be larger or smaller than the classical model
\citep{Bitsch2015, Suzuki2016}. In this paper, we mainly forced on
the influence of the giant planet on the final configuration of the
inner terrestrial planets. Considering that Kepler-68 is a
solar-mass star, we regard as MMSN the power-law index of the gas
disk density $k=3/2$.  $t_n$ is the disk depletion timescale, which
is observed to be approximately few million years
\citep{Haisch2011}. In this work, we assume $t_n=10^6$ yr. In the
simulation, an inner hole of the gas disk occurs around the central
star due to the star magnetic field, thus the gas disk is roughly
truncated at the corotation radius of star about nine stellar radii
\citep{Koenigl1991}. The stellar radius would be 2-3 times larger
before the protostar becomes a main sequence object. The radius of
central star for Kepler-68 system is approximately 1.24 $R_\odot$
\citep{Batalha2013}, consequently the inner boundary of the gas disk
is at roughly 0.1-0.15 AU. With the evolution of central star, the
truncation radius will decrease. It is possible that the truncation
radius is smaller than 0.1-0.15 AU in the planet formation stage
\citep{Bouvier2014}. Under such estimation, we set the inner edge of
the gas cavity to be 0.1 AU.

\subsection{Type I migration and eccentricity damping} \label{subsec:type1}
For a planet embedded in a protoplanetary disk, the exchange of
angular momentum between planet and gaseous disk will trigger
orbital migration of planets. If the mass of the planet is not
larger than $30~M_\oplus$, the variation of angular momentum on
planets will give rise to type I migration \citep{Tanaka2002,
2000MNRAS.315..823P}, the timescale of Type I migration on planet
with a mass $m$ is given by

\begin{equation}\label{equ:migI}
\begin{aligned}
\tau_{\rm migI}=&\frac{a}{|\dot{a}|}=\frac{1}{f_1}\tau_{\rm linear} \\
=&\frac{1}{f_1(2.7+1.1\beta)}\left(\frac{M_{\ast}}{m}\right)\left(\frac{M_{\ast}}{\Sigma_g a^2}\right)\left(\frac{h}{a}\right)^2\\
&\times
\left[\frac{1+(\frac{er}{1.3h})^5}{1-(\frac{er}{1.1h})^4}\right]\Omega^{-1}
\ yr,
\end{aligned}
\end{equation}
where $\tau_{\rm linear}$ is the linear analysis result and $f_1$ is
a reduction factor of migration speed. In our simulations, we
suppose $f_1$ ranging from 0.1 to 1. $\beta$ is the coefficient and
$\beta = -\mathrm{d} {ln \Sigma_g} / \mathrm{d} ln a$. $M_*$ is the
mass of central star and $\Omega=\sqrt{GM_{\ast}/r^3}$ presents
Keplerian angular velocity. $h$, $r$, $a$, $e$ and $m$ are scale
height of disk, distance between planet and star, semi-major axis,
eccentricity and mass of planet, respectively. Considering the disk
model we used, the timescale of Type II migration is much longer
than that of Type I migration \citep{Duffell2014, Durmann2015}, the
movement of giant planet could be negligible \citep{Baruteau2014}.
Here we ignore Type II migration of Planet d.

In addition, planet-disk interaction results in damping of orbital
eccentricity over a timescale of \citep{Cresswell2006}
\begin{equation} \label{equ:edamp}
\begin{aligned}
\tau_{\rm e}=&\left(\frac{e}{\dot{e}}\right)_{\rm edamp} \\
=&\frac{Q_e}{0.78}\left(\frac{M_{\ast}}{m}\right)\left(\frac{M_{\ast}}{a^2\Sigma_g}\right)\left(\frac{h}{r}\right)^4\Omega^{-1}\left[1+\frac{1}{4}\left(e\frac{r}{h}\right)^3\right]
\ yr,
\end{aligned}
\end{equation}
where $Q_{\rm e}$ is a normalization factor and here we adopt
$Q_{\rm e}=0.1$. Other symbols are the same as in Equation
\ref{equ:migI}. Orbital migration and eccentricity damping, induced
by the gas disk, will diminish when the planet enters into the inner
hole.

\section{Numerical simulation results} \label{sec:results}
To explore the configuration formation of Kepler-68 planetary
system, we assume two terrestrial planets are originally born far
away from their nominal locations, and then undergo type I migration
caused by the gas disk, along with a settled outermost gas-giant.
Therefore, the acceleration of the terrestrial planet with $m_i$ is
expressed as

\begin{eqnarray} \label{equ:acc}
\frac{d}{dt}\textbf{V}_i =
 -\frac{G(M_*+m_i
)}{{r_i}^2}\left(\frac{\textbf{r}_i}{r_i}\right) +\sum _{j\neq i}^N
Gm_j \left[\frac{(\textbf{r}_j-\textbf{r}_i
)}{|\textbf{r}_j-\textbf{r}_i|^3}- \frac{\textbf{r}_j}{r_j^3}\right]
\nonumber\\
+\textbf{F}_{\rm damp}+\textbf{F}_{\rm
migI},~~~~~~~~~~~~~~~~~\label{eqf}
\end{eqnarray}
where
\begin{eqnarray}
\begin{array}{lll}
\textbf{F}_{\rm damp} = -2\frac{\displaystyle (\textbf{V}_i \cdot
\textbf{r}_i)\textbf{r}_i}{\displaystyle r_i^2\tau_{\rm e}},
\\
\cr\noalign{\vskip 0.5 mm} \textbf{F}_{\rm
migI}=-\frac{\displaystyle \textbf{V}_i}{\displaystyle 2\tau_{\rm
migI}}. \label{dm}
\end{array}
\end{eqnarray}

For the gas-giant, we only consider the gravitational interaction
with other planets and the central star. To simulate the orbital
evolution of each planet, we have modified N-body integrator of
MERCURY6 \citep{Chambers1999}. In our simulations, the time step for
each integration is set to be less than 1/50 of orbital period of
the innermost planet and the integration accuracy is $10^{-15}$. The
argument of pericenter, longitude of ascending node and mean anomaly
(hereafter the three angles) of planets are randomly generated
between 0\degr and 360\degr. The eccentricities are initially
assumed to be zero and all planets are supposed to be co-planar.

In this work, we have performed 1610 runs in total, with an
integration timescale ranging from 0.25-2 Myrs, depending on the
system stability and planetary migration rate. It is possible that
there are more than two terrestrial planets in the systems. In this
paper, we mainly focused on studying the influence on the
configuration between the confirmed planets. Considering other
possible existed terrestrial planets is not massive enough to
influence the final configuration of planet b and c, we only assume
that there are two terrestrial planets in the system. \S
\ref{subsec:migration} shows the results of the the systems composed
of only two terrestrial planets for Group 1, whereas in \S
\ref{subsec:perturbation}, we investigate the formation of the inner
terrestrial planets affected by the giant planet for Group 2. For
all runs, we label the innermost planet as $P_b$ and the middle
planet as $P_c$, where the subscripts $b$ and $c$ denote each of
them, respectively, while $P_d$ represents the outermost companion
in the system.

\begin{figure*}
\centering
\includegraphics[width=18cm,height=14cm]{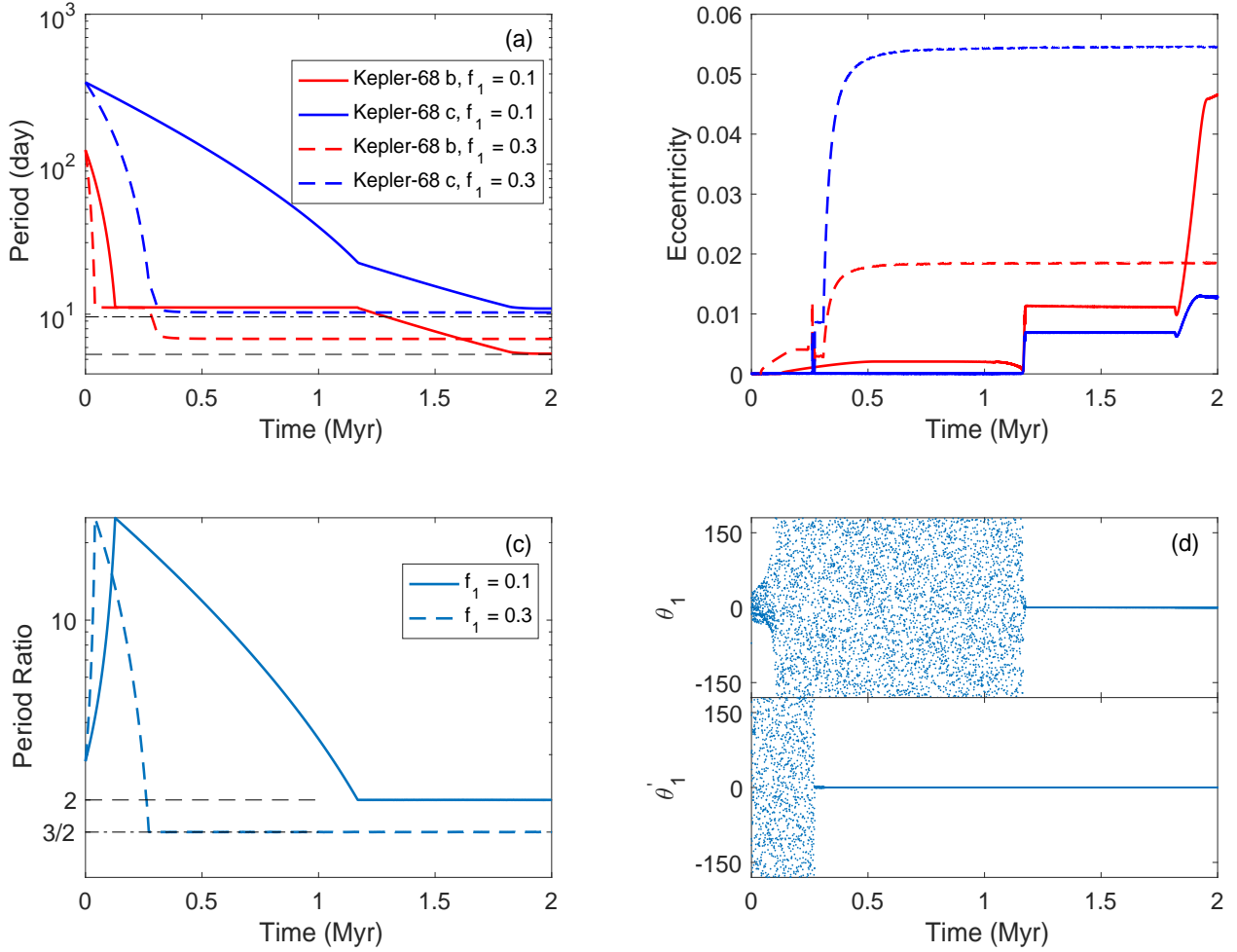}
\caption{Orbital evolution of G1-1 and G1-2. Solid lines represent
the run with $f_1=0.1$ (G1-1), whereas the dashed lines for
$f_1=0.3$ (G1-2). The initial locations of $P_b$ and $P_c$ are at
0.5 and 1 AU, respectively. Panel (a), (b), (c) and (d) show the
evolution of orbital period, eccentricity, period ratio, and
resonant angle, respectively. In Panel (a), the dashed and
dash-dotted lines (in gray), respectively, illustrate the currently
observed orbital period of Kepler-68 b and Kepler-68 c. Red and blue
lines of two upper Panels, respectively, display the orbital period
and eccentricity against the time over 2 Myrs for two planets. $P_b$
and $P_c$ are trapped in 2:1 MMR for $f_1=0.1$, while captured in
3:2 MMR for $f_1=0.3$. Panel (d) provides the time variation of the
resonant angles  $\theta_1=\lambda_1-2\lambda_2+{\varpi}_1$ of 2:1
MMR and  $\theta^{'}_{1}=2\lambda_1-3\lambda_2+{\varpi}_1$ of 3:2
MMR, which librate slightly about $0\degr$ when they enter into each
MMR. \label{fig:2pmig}}
\end{figure*}

\subsection{Group 1: Terrestrial planets migrate without giant planet} \label{subsec:migration}

In this Section, we carry out a series of simulations using a wide
variety of initial positions of $P_b$ and $P_c$, to examine
configuration formation for the terrestrial planet pair without the
presence of the giant planet in the Kepler-68 system. The
contribution of disk to planets is described as in Equation
(\ref{equ:migI}) and (\ref{equ:edamp}). We let $f_1=0.1, 0.3, 0.5,
0.8$ and $1$, respectively. Considering the currently observed
semi-major axis of $P_d$ and the inner edge of gaseous disk.
According to \citet{Kokubo98}, after oligarchic growth, the orbital
separations between planets are wider than 5 Hill radius, and the
typical orbital separation is about 10 Hill radius. Based on
\citet{Zhou2007}, the stable time for the planets with few
earth-mass is larger than $10^5$ years when the separation between
then is larger than 5 Hill radius. And according to the estimation
of \citet{Ford2001}, planetary system would maintain stable if their
relative orbital separation is larger than 3.5 Hill radius.
Therefore, in this work, we set the initial location of $P_b$
ranging from 0.2 to 1.2 AU, whereas that of $P_c$ spans from 0.3 to
1.3 AU with an equally-spaced separation of 0.1 AU which makes the
initial separation between two terrestrial planets larger than 3.5
Hill radius. Then we have performed 66 simulations for each $f_1$ by
means of the initials of $P_b$ and $P_c$. Thus we entirely implement
330 runs.

\begin{table*}
\small \centering \caption{Parameters adopted in our simulations for
G1 and G2.}
\begin{tabular*}{18cm}{@{\extracolsep{\fill}}cccccccccccc}
\hline \hline
 Case No.& $p_{b0}$&$p_{c0}$ &$f_1$&$p_{bf}$&$p_{cf}$&$e_{bf}$&$e_{cf}$&Giant Planet&$p_{d0}$&$e_{d0}$&$m_d$\\
&(day)& (day) &&(day)&(day)&&&&(day)& &$(M_J)$\\
\hline
G1-1&124&352&0.1&5.453&10.91&0.047&0.013&N&-&-&-\\
G1-2&124&352&0.3&6.844&10.27&0.019&0.055&N&-&-&-\\
G2-1&124&352&0.1&5.448&10.9&0.053&0.012&Y&580&0.0&0.97\\
G2-2&124&352&0.3&6.844&10.27&0.019&0.055&Y&580&0.0&0.97\\
G2-3&312&261&0.8&6.844&10.27&0.055&0.019&Y&580&0.0&0.97\\
G2-4&124&352&0.3&6.17&10.28&0.0165&0.053&Y&580&0.18&0.97\\
G2-5&33&60&0.1&6.111&10.69&0.033$\pm$ 0.0046&0.0138$\pm$ 0.013&Y&129&0.03&10.0\\

\hline \hline \label{allcase}
\end{tabular*}

\begin{tablenotes}
    \footnotesize
    \item[]Note: $p_{b0}$ and $p_{c0}$, respectively, represent the initial orbital period of $P_b$ and $P_c$, while $p_{bf}$ and $p_{cf}$ stand for each of the final period. $e_{bf}$ and $e_{cf}$ denote each of the final eccentricity. $p_{d0}$, $e_{d0}$, and $m_d$ represent the initial orbital period, eccentricity, and mass of $P_d$.
    \end{tablenotes}

\end{table*}

From the numerical outcomes, we find that the initial position of
planets and their mutual separation actually have no remarkable
influence on the final configuration of the planetary system.
However, the speed of type I migration plays a key role in affecting
the final configuration of the planet pair. For $0.3 \leq f_1 \leq
1$, Kepler-68 b and c prefer to be captured into 3:2 MMR, whereas
for $f_1=0.1$, planets with lower migration speed are trapped into
2:1 MMR rather than 3:2 MMR. These results are consistent with those
of \citet{Wang2012}. The  typical evolution for the cases is shown
in Figure \ref{fig:2pmig} for $f_1=0.1$ and $f_1=0.3$, respectively.
The major initial conditions and the final results are shown in G1-1
and G1-2 of Table \ref{allcase}. Planet b and c are assumed to start
migrating from 0.5 and 1 AU for two cases, respectively. Solid lines
illustrate the results of $f_1=0.1$, while the dashed blue and red
lines display those outcomes of $f_1=0.3$. The gray lines indicate
the nominal locations of two terrestrial planets in the system. As
shown in Figure \ref{fig:2pmig}, without a remarkable reduction in
the speed of migration, the terrestrial planets are quickly locked
into 3:2 MMR at about 0.03 Myr, while for a lower migration speed,
the planets are trapped into 2:1 MMR at about 1.2 Myr. The final
eccentricity of planet c is a bit higher than planet b in G1-2, as
compared with that of G1-1.

As planet b is more massive than planet c, $P_b$ always migrates
faster than $P_c$ according to the timescale of type I migration
given in Equation (\ref{equ:migI}) before they arrive at the inner
edge of gas disk. Once $P_b$ reaches the inner boundary of gas disk,
the planet will halt migrating because of the absence of gas. When
$P_c$ approaches $P_b$, the two terrestrial planets will be captured
into resonance and the planet pair will migrate in the same pattern
until $P_c$ stops migration near the inner edge of gas disk. Based
on our estimation, the inner boundary of the gas disk locates at
about 0.1 AU, which is very close to the observed orbit of $P_c$.
Thus, through type I migration from outside, $P_c$ can be formed
near its nominal location. If the inner
boundary of gas disk locates closer than 0.1 AU \citep{Bouvier2014},
the final location of $P_c$ may change as the inner disk edge
moves much closer to the star. Therefore, the location of disk edge
suggests that terrestrial planets may have been born at the early stage
of the central star that the truncation radius is around 0.1 AU according to our estimation.

In our simulations, we find that 80\% of planet pairs are trapped in
3:2 MMRs, whereas 20\% of planet pairs are involved in 2:1 MMRs at
the end of runs. The simulations indicate that the planet pairs are
entirely trapped into first-order MMRs concerning with a wide
variety of initial region and speed of type I migration. The
resultant orbital periods of $P_c$ is approximately 10 days, while
that of $P_b$ is observed to be either 5.45 or 6.84 days, depending
on what kinds of MMRs they enter into. If
there are additional terrestrial planets migrating toward the
proximity of the inner two terrestrial planets, the final
configuration of inner planet pair is similar to the results as shown
in Group 1, but it will increase possibility of planet pair involved in
more compact configuration. For the systems that are composed of three terrestrial planets, the planet pairs have more opportunity to be in 3:2 MMR
\citep{Wang2014, Wang2017} in the migration scenario. Therefore, here our primary goal is to investigate how the configuration of inner planet pair is affected by a distant gas-giant in the system for only two terrestrial planets formed.

Here we choose G1-1 and G1-2 as two typical models to examine the
situation of formation of inner planet pair affected by the
outermost giant planet. Using similar initial parameters of G1-1 and
G1-2, we then carry out 100 simulations for each $f_1$. The
simulations reveal that the planet pairs are entirely trapped into
2:1 MMR for $f_1=0.1$, whereas they are associated with 3:2 MMR for
$f_1=0.3$. These results agree with those of G1-1 and G1-2.
According to equation (2), the timescale of type I migration is
proportional to $(mr)^{-1}$. The masses of inner terrestrial planets
in the system is 5.98 and 2.18 $M_\oplus$. Based on these initial
settings, the speed of type I migration of the innermost planet in
higher than the second terrestrial planet initially. The innermost
planet migrate to the inner edge first and then the second one catch
up with the innermost planet around the disk edge. Therefore, the
relative speed between two terrestrial planets is not so high
leading to the capture of 2:1 and 3:2 MMRs rather than 4:3 MMR
\citep{Mustill2011, Ogihara2013}.

\begin{figure*}
\centering
\includegraphics[width=18cm,height=14cm]{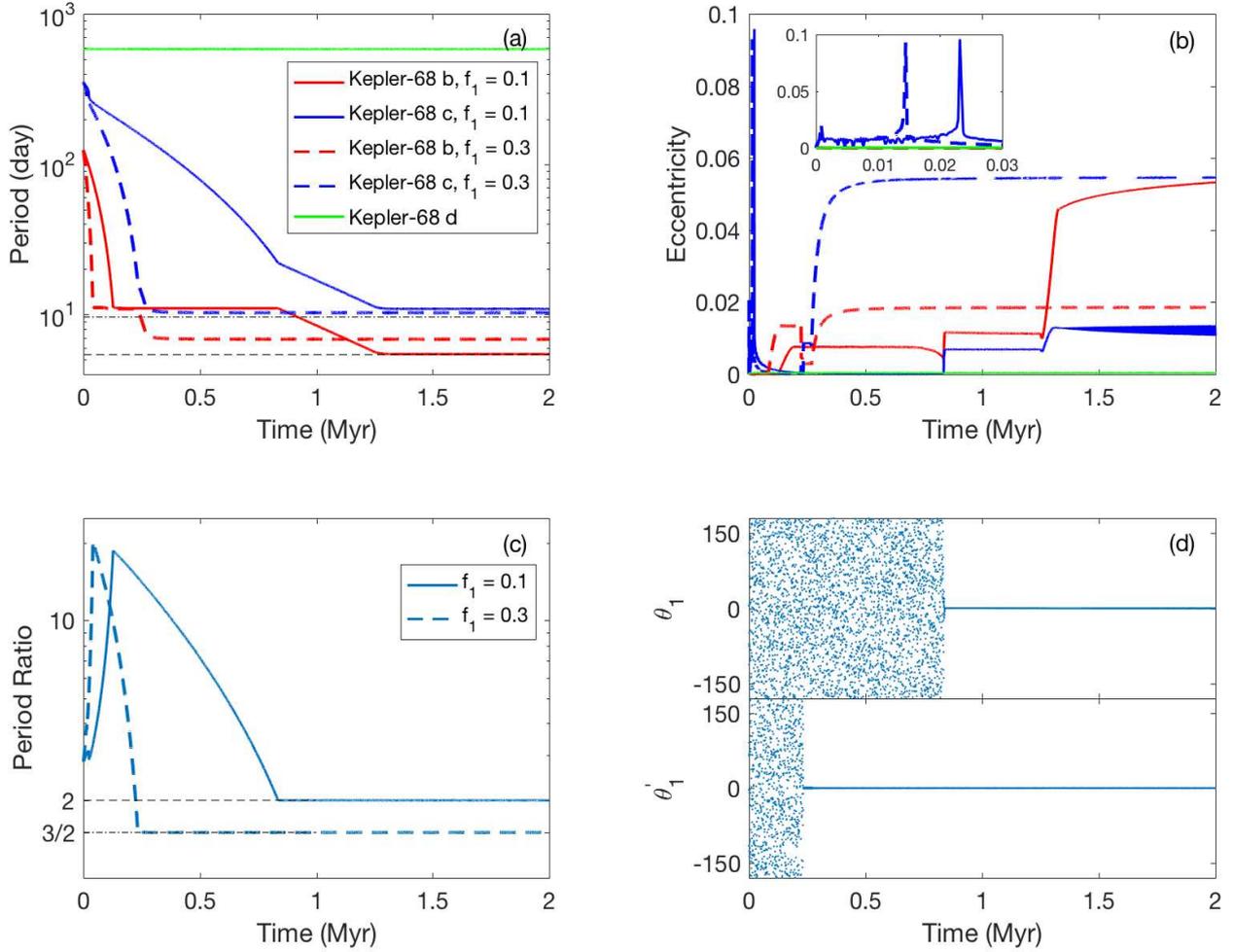}
\caption{Same as in Figure \ref{fig:2pmig}, orbital evolution of
G2-1 and G2-2. A giant planet d (by green line) initially orbit at
1.4 AU with $e_d=0$. Panel (b) exhibits that the eccentricity of
Kepler-68 c is excited by the giant planet at the early stage of
migration but damped rapidly due to the disk. The giant planet can
speed up the migration of two inner terrestrial planets. Panel (d)
shows the resonant angles $\theta_1=\lambda_1-2\lambda_2+{\varpi}_1$
of 2:1 MMR and  $\theta^{'}_{1}=2\lambda_1-3\lambda_2+{\varpi}_1$ of
3:2 MMR fluctuate slightly about $0\degr$ when two inner terrestrial
planets are captured into each MMR. \label{fig:3pmig}}
\end{figure*}

\subsection{Group 2: Terrestrial planets migrate with perturbation of giant planet} \label{subsec:perturbation}

In Kepler-68 system, \citet{Gilliland2013} showed that the
additional giant planet locates at about 1.4 AU, which is far away
from the nominal regime  of terrestrial planets. In such
circumstance, the formation of inner terrestrial planets $P_b$ and
$P_c$ would be influenced due to the existence of the gas-giant.
Some study suggests that the giant planets can be formed in the
system earlier than the terrestrial planets \citep{Lykawka2013}.
Thus, in Group 2, we assume the giant planet $P_d$  first  occurred
in the outer region compared with the two inner companions, and we
will extensively investigate the migration of the inner planets
induced by the perturbation of giant planet in the system. To
further clarify this issue, here we perform more simulations of
three subgroups based on the orbital period and eccentricity of
$P_d$.  In the following, we will present the detailed exploration.

Subgroup 1: $P_d$ initially locates on circular orbit at 1.4 AU.
Same as in Group 1, 330 runs are performed using various initial
locations of two terrestrial planets, and other 200 cases are run
for planet b and c at 0.5 and 1 AU, respectively, for $f_1=0.1$  and
$f_1=0.3$, respectively.

Subgroup 2: $P_d$ originally moves on an eccentric orbit of
$e_d=0.18$  at 1.4 AU, whereas $P_b$ and $P_c$ reside at 0.5 and 1.0
AU, respectively, at the beginning of simulations. Here we adopt
$f_1=0.1$ and $f_1=0.3$. And we run 200 simulations.

Subgroup 3: $P_b$ and $P_c$ locate at 0.2 and 0.3 AU initially. We
adopt $f_1=0.1$ and $f_1=0.3$, respectively. To explore the
formation of two terrestrial planets influenced by the gas-giant on
circular orbits, we perform 150 runs with a family of combined
parameter of $m_d$ of 1, 5, and 10 $M_J$, $a_d$ ranging from 0.4 to
0.8 AU. As a comparison, we then carry out 200 additional runs in
the case that $P_d$ moves on eccentric orbits at 0.5 AU for
$e_d=0.03$ and $e_d=0.1$, respectively.

\subsubsection{Subgroup 1: Giant planet on circular orbit formed at its nominal location}
First of all, we consider similar cases as in \S
\ref{subsec:migration}, but with an additional giant planet $P_d$
lying beyond $P_b$ and $P_c$ on a circular orbit. The initial mass
and semi-major axis of $P_d$  in the simulations are the observed
values as given in Table \ref{tab:k68}.

Among 330 cases,  24\% of unstable systems in the results are
induced by the small separation which is near the boundary of
$3.5~R_H$. In this section, we mainly discuss the results of 250
stable cases.

Figure \ref{fig:3pmig} shows the dynamical evolution of two typical
stable cases. To compare with the results in Group 1, we adopt the
same initial parameters of G2-1 and G2-2 as those of G1-1 and G1-2.
However, to better understand how the configuration of two
terrestrial planets is affected, in the simulations we add the
perturbation arising from the giant planet with a circular orbit.
Panel (b) of Figure \ref{fig:3pmig} exhibits the eccentricity of
$P_c$ is excited to about 0.1 at the very beginning . However, as
the planet interacts with the gas disk, the excited eccentricity can
go down to near zero very quickly. In the meantime, the semi-major
axis of planets decreases due to angular momentum conservation.
Clearly, it is easy to note that the migration speed of $P_c$ in
G2-1 and G2-2 is faster than that in Group 1. Panel (d) shows the
resonant angles $\theta_1=\lambda_1-2\lambda_2+{\varpi}_1$ of 2:1
MMR and  $\theta^{'}_{1}=2\lambda_1-3\lambda_2+{\varpi}_1$ of 3:2
MMR librate slightly about $0\degr$ when two inner terrestrial
planets are captured into each MMR. Although there exists a giant
planet, the configuration of the terrestrial planet pair is not
changed for this group.

\begin{figure*}
\centering
\includegraphics[width=18cm,height=7cm]{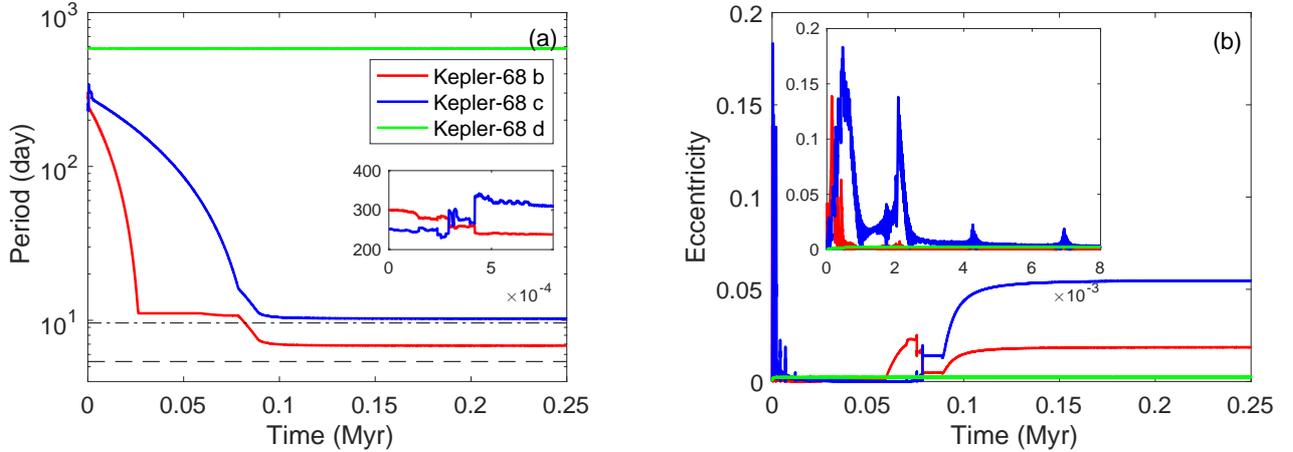} 
\caption{The evolution of G2-3. Panel (a) and (b) show the evolution
of orbital periods and eccentricities, respectively. The zoomed
window in Panel (a) displays that the two planets exchange their
orbits at a very early stage with respect to the excitation of
eccentricity in Panel (b). Note that the inner terrestrial pair can
be trapped into 3:2 MMR in the evolution. \label{fig:exchange}}
\end{figure*}

When planets move inward, the eccentricities of $P_b$ and $P_c$ can
be excited for several times as they approach to the gas-giant. In
the evolution, terrestrial planets have opportunities to exchange
their orbits. The excitation always happens at the earlier evolution
of simulations, the gas disk still remains dense enough to damp the
eccentricity. Thus, the system can retain stable after the orbital
exchange. Here we come to conclusion that the systems like Kepler-68
might have undergone orbital exchange. According to the estimation
of isolation mass, the mass of solid core is proportional to
$a^{3/4}$ \citep{Ida2004}. In Kepler-68 system, $P_b$ is more
massive than $P_c$. They may exchange their orbits once. Figure
\ref{fig:exchange} shows the evolution of a typical case that $P_b$
and $P_c$ locate at 0.9 and 0.8 AU, respectively. The starting
conditions are denoted by G2-3 in Table \ref{allcase}. They migrate
with $f_1=0.8$. $P_d$, moving on circular orbit, originally resides
at 1.4 AU. The eccentricity of $P_b$ and $P_c$ can be frequently
excited to about 0.1 and decline to near zero within a shorter
timescale because of the disk. Orbital exchange occurs at about a
few hundred years. The subsequent evolution of two terrestrial
planets bears a resemblance to that of G2-1 and G2-2. Consequently,
the inner planet pair is trapped into 3:2 MMR, which coincides with
that of G2-2. Although the orbital exchange happened in case G2-3,
there is only one case in this group, the possibility is about
0.4\%. Additionally, the initial orbital
separation of $P_b$ and $P_c$ is $\sim$ 6 $R_H$. As the growth is
faster for the inner planet, it is likely that the inner planet
starts inward migration before the outer planet completes the
growth. The orbital separation between two terrestrial planets tends
to be larger. Considering the initial conditions and the possibility
in this case, such orbital exchange may hardly occur, except for some particular conditions.

With a giant planet on circular orbit at its nominal location, if
terrestrial planet moves so close to $P_d$, the separation between
them satisfies $\Delta \leq 0.2 \rm ~AU$, then the system becomes
unstable. If the terrestrial planet is close to $P_d$ and suffer
fast migration, orbital exchange can take place between two inner
planets. Although the gas-giant can speed up migration rate of two
terrestrial planets, it has no direct influence on resultant
configuration of inner planet pair. Furthermore, we conduct 200 runs
to examine this  using similar parameters like those of G2-1 and
G2-2, along with the variable three angles of planets. We find that
the outcomes are consistent with those of G2-1 and G2-2.

\begin{figure*}
\centering
\includegraphics[width=18cm,height=14cm]{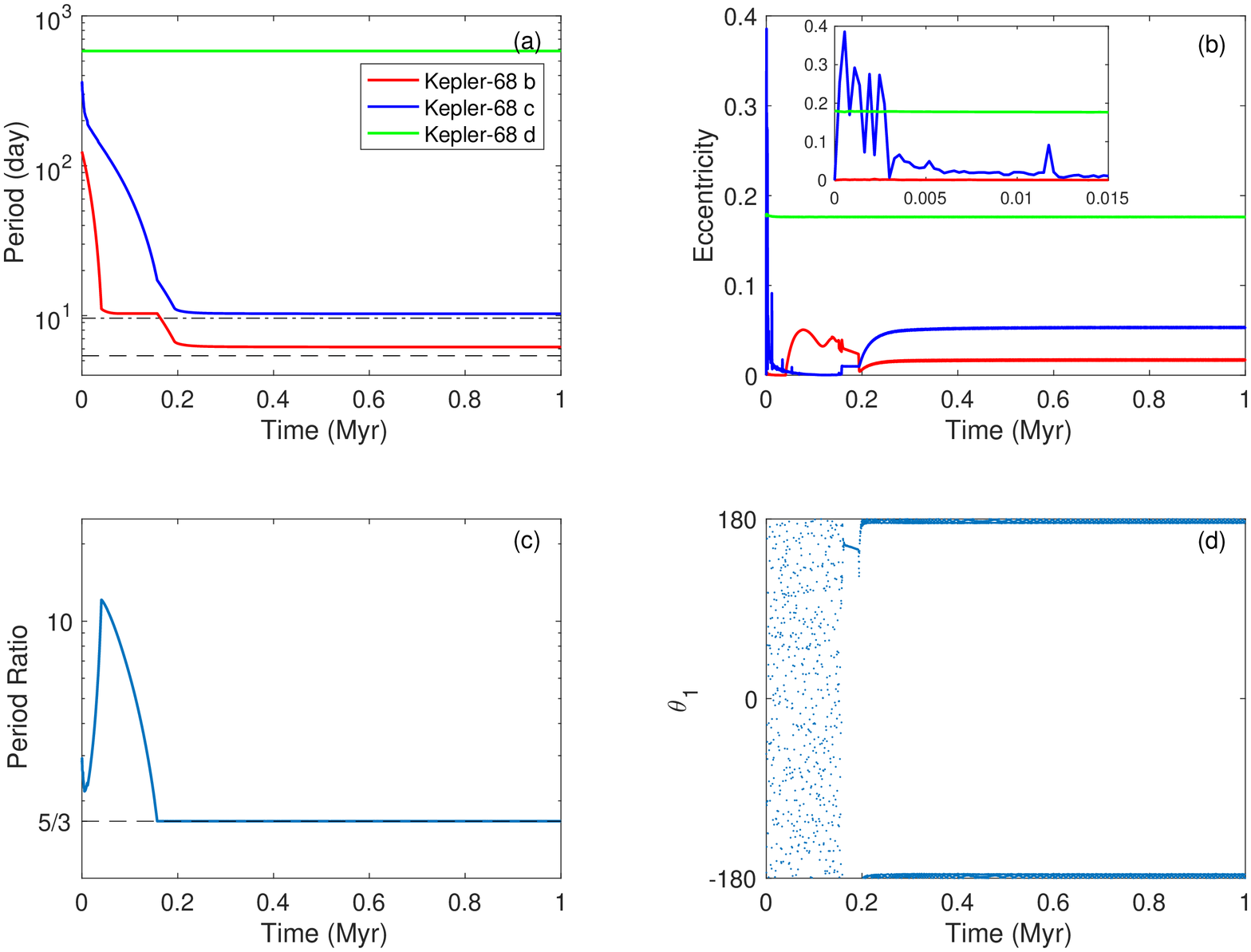}
\caption{Orbital evolution of G2-4. The label of each panel and
colored lines are the same as in Figure \ref{fig:2pmig}. A giant
planet marked by  green line locates at 1.4 AU with an initial
eccentricity 0.18. Two inner terrestrial planets can be captured
into 5:3 MMR at about 0.18 Myr. The resonant angle
$\theta_1=3\lambda_1-5\lambda_2+2{\varpi}_1$ for 5:3 MMR librates
about $180\degr$ with a tiny amplitude as shown in Panel (d).
\label{fig:5v3mmr}}
\end{figure*}

\subsubsection{Subgroup 2: Giant planet on eccentric orbit formed at its nominal location}
As shown in Table \ref{tab:k68}, $P_d$ is reported to be on
eccentric orbit with $e_d=0.18\pm 0.05$.It is possible that $P_d$
has already got its nominal eccentricity before the formation
process of the inner terrestrial planets. Here we further study the
configuration formation of inner terrestrial pair due to the
perturbation of a giant planet on eccentric orbit. Here $P_d$ is
assumed to be 1.4 AU with an initial eccentricity $e_d=0.18$ . The
locations and eccentricities of $P_b$ and $P_c$, and $f_1$ are
chosen as those of G1-1 and G1-2. Thus we carry out 100 cases for
$f_1=0.1$ and $f_1=0.3$, respectively.

The results show that $\sim$ 83\% and $\sim$ 82\% of the systems are
unstable for $f_1=0.1$ and $f_1=0.3$, respectively. Here we are
particularly interested in those of stable cases involved in MMRs.
For $f_1=0.1$, there are 4\% and 1\% of the systems hosting two
terrestrial planets locked in 2:1 and 3:1 MMR, respectively. By
contrast, 4\%, 10\% and 4\% separately harbor the inner planet pair
in 2:1, 3:2 and 5:3 MMRs for $f_1=0.3$.

Figure \ref{fig:5v3mmr} illustrates the evolution of a typical run
with two terrestrial planets in 5:3 MMR at the end of simulation.
The initial conditions and final locations are displayed as
{\color{blue}{\textbf{G2-4}}} in Table \ref{allcase}. Here the
eccentricity $e_c$ of the middle planet is able to be excited up to
0.4 while $e_b$ still remains a low value. In the simulations, both
of $P_c$ and $P_d$ sustain eccentric orbits that may give rise to
multiple  orbit-crossing between them, thereby stirring $e_c$
repetitively. From Panel (b) in Figure \ref{fig:5v3mmr}, we can see
that $e_c$ can be pumped up to above 0.3 within several thousand
years. Subsequently, the migration speed of $P_c$ increases all the
time as a result of the damping of eccentricity induced by the gas
disk, hence $P_c$ arrives at the inner edge of gas disk in advance.
However, we should emphasize that the inner planet pair can depart
from 2:1 or 3:2 MMRs as $P_c$ speeds up. As can be seen by Panel (d)
, $P_c$ and $P_b$ are tuned into 5:3 MMR, where the resonance angle
$\theta_1=3\lambda_1-5\lambda_2+2{\varpi}_1$ librates about
$180\degr$ for  1 Myr. This reminds us a very close configuration
compared to the observed Kepler-68 system as reported in Table
\ref{tab:k68}.

With a giant planet on eccentric orbit at its nominal location, a
similar configuration can be produced through our formation
scenario. If terrestrial planet formed at the outer region, which is
quite close to the nominal location of giant planet, the eccentric
orbit of giant planet will have influence on the nearby terrestrial
planet impulsively leading to speeding up of the adjacent planet. In
summary, we conclude that the inner terrestrial planet pair can be
trapped into second-order MMR, which  differs from the simulation
results without a giant planet in the outer region.

\begin{figure*}
\centering
\includegraphics[width=18cm,height=14cm]{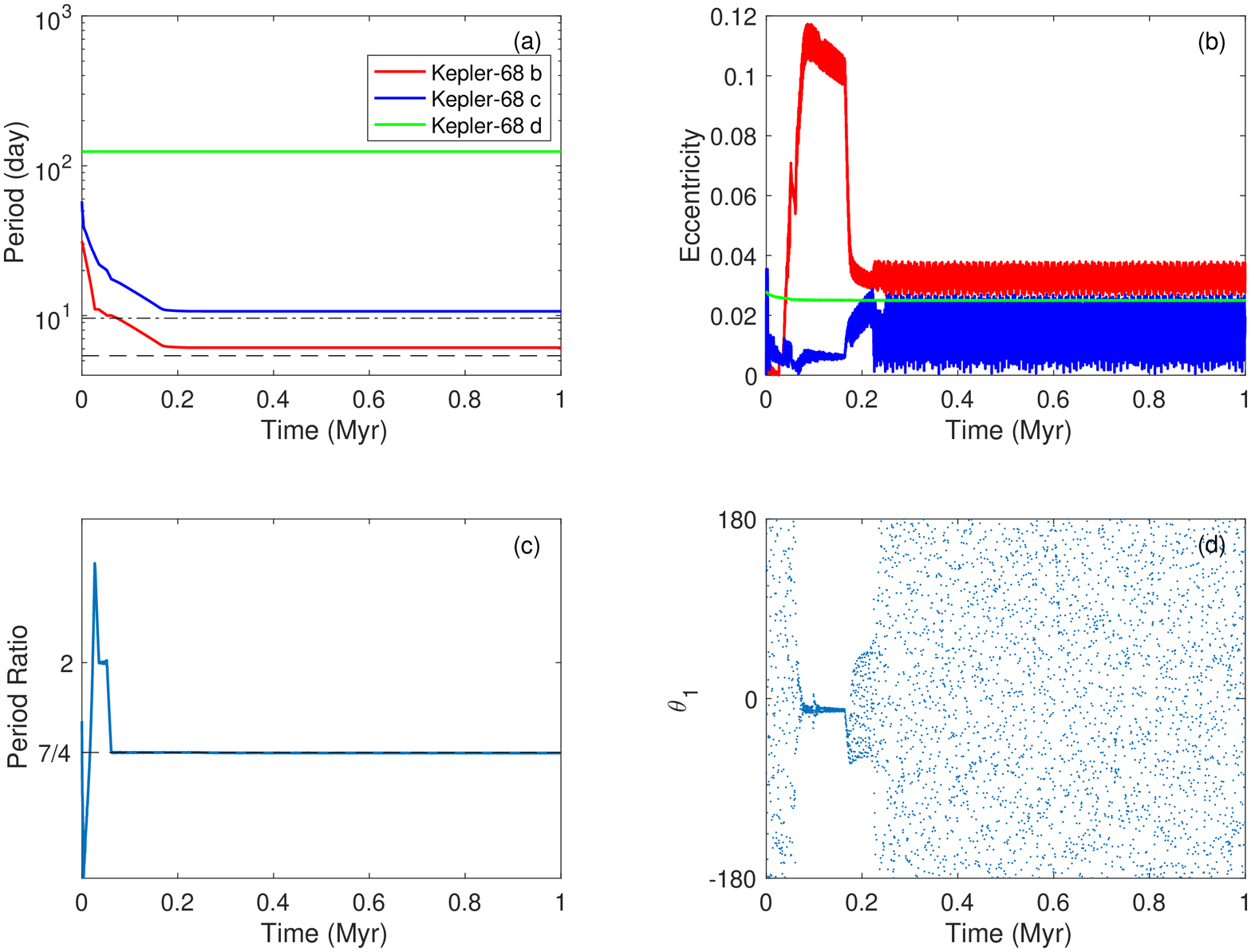}
\caption{Same as in Figure \ref{fig:2pmig}, orbital evolution of
G2-1 and G2-2. The initial locations of $P_b$, $P_c$ and $P_d$ are
0.2, 0.3 and 0.5 AU, respectively. Here $e_d$ is 0.03. Note that
$P_b$ and $P_c$ are temporarily trapped into 2:1 MMR at about 0.035
Myr, subsequently they have been involved in 7:4 MMR since 0.077 Myr
along with a resonant angle
$\theta_1=4\lambda_1-7\lambda_2+3{\varpi}_1$ tuning libration into
circulation. \label{fig:gf03}}
\end{figure*}

\subsubsection{Subgroup 3: Giant planet on eccentric orbit with different masses and semi-major axes}
As aforementioned, the existence of $P_d$ can affect the final
configuration of the inner pair. Here we primarily aim to study the
circumstance for two terrestrial planets when $P_d$ has a diverse
mass and initial location.

First, we investigate dynamical evolution of the systems that harbor
$P_d$ on a circular orbit. In our simulations, we consider the
initial parameters of the giant planet with a combined parameter of
mass ($m_d=$ 1, 5, and 10 $M_J$), semi-major axes ($a_d=$ 0.4, 0.5,
0.6, 0.7, and 0.8 AU), and the speed of type I migration ($f_1=$ 0.1
and 0.3), then we perform 150 runs to explore how these parameters
play a part in the evolution of the planets. The results reveal that
the inner pair can be captured into 2:1 MMR for $f_1=0.1$ and 3:2
MMR for $f_1=0.3$, which is similar to those in Group 1.

Second, for $P_d$ on eccentric orbit at 0.5 AU, $P_b$ and $P_c$
occupying the orbits at $a_b=0.2$ and $a_c=0.3$ AU, respectively, we
carry out 200 simulations with respect to $e_d=0.03$ and $e_d=0.1$
for $f_1=0.1$. As mentioned above, the inner planet pair is involved
in 2:1 MMR for $f_1=0.1$ in the case of without a gas-giant or with
a giant planet on circular orbit. As a comparison, there is less
than 15\% of the runs trapped in 2:1 MMR for terrestrial planets.
Moreover, approximately 50\% of them turns to other MMRs.

For $e_d=0.03$, we conduct 100 simulations. From the results, we
find that 31\% of the runs occupy two terrestrial planets involved
in 3:2 MMR, whereas 12\%, 5\%, 2\% and 1\% harbor inner pairs in
5:3, 7:4, 7:5 and 8:5 MMRs, respectively. For $e_d=0.1$, 12\% hosts
the pairs in 3:2 MMR, whereas 24\% and 14\% are captured in 4:3 and
7:5 MMRs.

Figure \ref{fig:gf03} shows the dynamical evolution for a typical
case, which corresponds to two terrestrial planets in 7:4 MMR at the
end of the simulation. The initial conditions and final locations
are given in Table \ref{allcase}. As can be seen Figure
\ref{fig:gf03}, two migrating planets are temporarily trapped into
2:1 MMR at about 0.035 Myr, induced by a massive gas-giant of 10
$M_J$  at 0.5 AU. However, the terrestrial planets simply remain at
2:1 MMR for a very short time. $P_b$ and $P_c$ will escape from 2:1
to 7:4 MMR because of strong perturbation from the giant planet,
providing the evidence of formation of Kepler-68 system. In the
subsequent evolution, the eccentricity of $P_b$ could be stirred
dramatically up to about 0.12 when two planets enter into MMR.

For the simulations in relation to a more massive giant planet, we
find that the resultant configuration of the inner pair cannot be
altered when the gas-giant revolves around the host star in a
circular trajectory or move on an eccentric orbit of $e_d\leq 0.02$.
However, in the case of the giant planet orbiting its central star
with $e_d>0.02$, the eccentricity of inner terrestrial especially
the one closer to the giant planet will be excited to higher than
0.1 at least once. In some cases, the eccentricity of $P_c$ can be
excited to be about 0.4 for several times. Due to the eccentricity
damping caused by the gas disk, the orbital migration will be
speeded up. Meanwhile, the resonance web is dense for compact
planetary systems \citep{2015MNRAS.446.1998N}. Thus, inner
terrestrial planet pair have chance to captured into high order
MMRs.These outcomes favor formation scenario for two inner planets
due to a distant gas-giant with an eccentric orbit.

\subsection{Group 3: Secular Resonance} \label{sec:secular}
The gas disk inside the orbit of giant planet might have been
seriously depleted by planetary or stellar accretion
\citep{Nagasawa2005} when  terrestrial planets migrate to the region
near its observed orbit. Then gaps are created around orbits of
planets \citep{Goldreich1980, Takeuchi1996}. The configuration of
inner planet pair will be reshaped by secular perturbation stemming
from the outermost companion and the gas disk exterior to planets.

The gravity of disk with an inner boundary at $d$ ($d > a_d$)
contributes to the planet $i$ is given by the formula
\citep{Nagasawa2003}
\begin{equation} \label{equ:sr}
{\bf f}_{i,disk}=8 \pi G \Sigma_g\left(r_i\right)\frac{\bf r_i}{r_i}
\displaystyle\sum_{n=0}\left[B_n
\left(\frac{r_i}{d}\right)^{2n+1/2}\right],
\end{equation}
where
\begin{equation}
B_n \equiv \left[\frac{(2n)!}{2^{2n}\left(n!\right)^2}\right]^2
\frac{n}{\left(4n+1\right)}.
\end{equation}
$\Sigma_g(r)$ refers to the same as in Equation (\ref{equ:sigma}).
Here we set $d=2 ~\rm AU$. As a consequence, the eccentricity
damping and type I migration caused by the gas disk on the planets
disappear. The dissipation of disk can trigger secular perturbation
between planets, thereby leading to angular momentum exchange and
modifying their eccentricities \citep{Nagasawa2003}. The secular
resonances occur only when the mass of outer disk is comparable to
that of the giant planet \citep{Nagasawa2005}. The region of secular
resonance gradually moves inward as gas density decreases. And we
adopt an initial surface density of disk by Equation
(\ref{equ:sigma}) with a disk depletion timescale $t_n = 10^6$ yr.
According to the analysis of secular resonance
\citep{Heppenheimer1980} and the disk model in this work, the
location of secular resonance decreases from 0.8 AU to 0.02 AU which
sweeps through the region of the terrestrial planets in the system.

\begin{figure}
\centering
\includegraphics[width=\columnwidth,height=6cm]{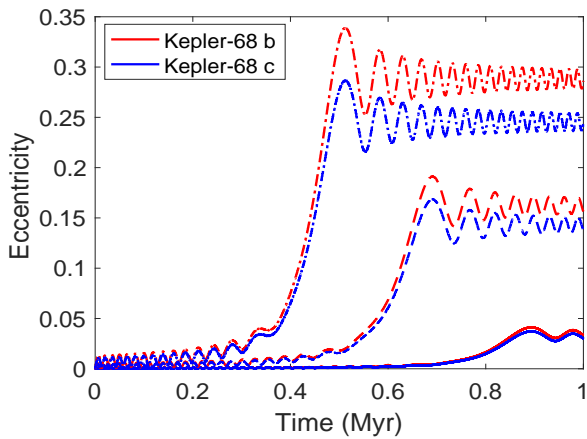}
\caption{Eccentricity evolution of Kepler-68 b and Kepler-68 c
(colored by red, blue) under secular resonance of the giant planet
and disk. The solid, dashed and dash-dotted lines each represent the
eccentricity evolution of the inner planet pair induced by a giant
planet with a mass of 1, 5 and 10 $ M_J$, respectively.
\label{fig:sr}}
\end{figure}

Figure \ref{fig:sr} shows the eccentricities of the inner planet
pair (each colored by red and blue line) of Kepler-68 system evolve
over a timescale of 1 Myr, resulting from secular resonance by the
giant planet with a variable mass. Here $P_b$ and $P_c$ are
initially assumed to locate on circular orbits, and their masses and
semi-major axes are listed in Table \ref{tab:k68}. Additionally,
$e_d$ corresponds to the observed value 0.18. The solid, dashed and
dash-dotted lines represent the eccentricity evolution of $P_b$ and
$P_c$ induced by the giant planet of a mass of 1, 5 and 10 $~ M_J $,
respectively. As can be seen from Figure \ref{fig:sr}, we can see
that the amplitude of eccentricity excitation of two terrestrial
planets increases as the mass of gas-giant goes up. To be more
specific, for $m_d=1~ M_J$, the simulations indicate that both of
eccentricities are not well excited below 0.05. However, the
eccentricities can be pumped up to 0.15 for $m_d=5~ M_J$, while they
reach about 0.30 for $m_d=10~ M_J$.

As reported in Table \ref{tab:k68}, the eccentricity of $P_c$ is not
well confirmed. One of the orbital fittings indicates that $e_b$ and
$e_c$ both are nearly zero \citep{Gilliland2013, Mills2019}, whereas
an alternative orbital solution from observations show that $e_c$ is
approximately 0.42 \citep{VanEylen2015}. According to our outcomes,
$e_c$ can be stirred up to above 0.2 only if $m_d$ is more massive
than 5 $M_J$. In such cases, $e_b$ and $e_c$ can be further excited
in the evolution. On the other hand, as $P_b$  moves much closer to
the host star than $P_c$,  thus the tidal effect by the central star
plays a vital part in $P_b$, thereby resulting in the damping of
eccentricity within $2.7\times 10^5~Q'$ years, which seems to be
much longer than the timescale of secular resonance as shown in
Figure \ref{fig:sr} \citep{Mardling2014, Zhou2008}. Here we suppose
the density of terrestrial planet is about $3\rm \ g \ cm^{-3}$,
where $Q'$ is the tidal dissipation factor. Hence we can estimate
the timescale of eccentricity damping of $P_c$ by tidal effect is
approximately $6.6\times 10^6~Q'$ years, implying that the
eccentricity of $P_c$ is difficult to be damped.

On the other hand, we note that the eccentricity of $P_c$ is
possibly excited to be above 0.2 when $m_d$ is larger than 5 $M_J$.
In contrast, $P_c$ will stay at circular orbit with a less massive
giant planet.

\section{Discussion and Conclusion} \label{sec:summary}
In this work, we primarily conduct thousands of numerical
simulations to explore the configuration formation for two inner
terrestrial planets near MMRs as a result of the perturbation of the
outermost gas-giant for the Kepler-68 system. Here we summarize the
major outcomes and conclude that,

\begin{enumerate}
\item
For the system only harboring two terrestrial planets, they are
inclined to be trapped into 2:1 or 3:2 MMRs, which depends on the
speed of type I migration. When $0.3\leq f_1 \leq 1$, the inner
planet pair prefers to be captured into 3:2 MMR, whereas with a low
speed of orbital migration $f_1 \leq 0.1$, the two terrestrial
planets are more likely to be associated with 2:1 MMR. The results
obtained here are consistent with those reported in our earlier work
\citep{Wang2014}.

\item
For the system composed of two terrestrial planets and an outermost
giant planet, we note that the inner planet pair can eventually
reach similar configuration in the secular evolution like those for
the two-planet system, when the giant planet orbits the central star
in a circular trajectory.  As a comparison, for the three-planet
system that one giant planet moves on eccentric orbit, we find that
the terrestrial planets can be involved in 5:3 MMRs, being
indicative of a close match with the currently observed orbits for
Kepler-68. Further study shows that the inner pair would be captured
in 7:4 or 8:5 MMRs if there is a more massive gas-giant at closer
orbit in the system.

\item
In view of the depletion of gas disk, the secular resonance may
sweep up the nominal locations of inner terrestrial planets. Hence,
the eccentricities of inner planets can be excited up to 0.2 with a
massive giant planet of $m_d\geq 5~M_J$. Otherwise, the terrestrial
planets retain to be circular orbits for a lower mass of the
outermost companion.
\end{enumerate}

\begin{figure}
\centering
\includegraphics[width=\columnwidth,height=6cm]{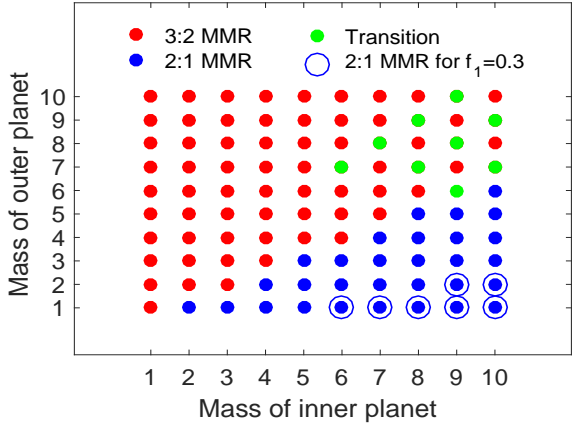}
\caption{Final configuration related to MMRs for two planets with a
mass ranging from 1 to 10 $M_\oplus$. Filled dots in red or blue
represent the systems involved in 3:2 and 2:1 MMRs for $f_1=0.1$,
respectively, while green dots stand for those of transition.
Circles indicate those of 2:1 MMR for $f_1=0.3$. \label{fig:mmr01}}
\end{figure}

Allowing for above-mentioned formation scenario, there are several
critical factors that may have influence on resultant configuration
of two terrestrial planets in the planetary system. First, the speed
of orbital migration can play a significant role in shaping the
final orbits of inner planet pair. With a faster orbital migration,
the planet pair prefers to habitat in a more compact configuration
\citep{Hands2016}. Second, the mass ratio, which is related to the
speed of orbital migration, can further affect the near-resonant
configuration for two terrestrial planets. To better understand this
circumstance, here we perform 2000 additional simulations to
investigate the systems by altering their initial masses in the
range of 1 to 10 $M_\oplus$ and the three angles. Figure
\ref{fig:mmr01} shows final configurations for two terrestrial
planets in MMRs. In our simulations, we assume the planets
originally locate at 0.2 and 0.3 AU, respectively. For $m_{\rm
b}/m_{\rm c}<1$, we observe that the two planets are inclined to be
involved in 3:2 MMR. Based on the estimation of isolation mass
\citep{Ida2004}, the mass of solid core corresponds to the
semi-major axis where they are formed with $m\propto a^{3/4}$.
Accordingly, the mass of terrestrial planet grows up as its
semi-major axis rises. If the system can maintain steady, the planet
pair probably tends to be trapped in 3:2 MMR, e.g., K2-19 and
Kepler-59 system \citep{Petigura2020, Saad2020}. Otherwise, they are
probably captured in 2:1 MMR. Third, the planet pair can be further
trapped in high-order MMRs owing to the presence of a massive giant
planet with an eccentric orbit. Additionally, the results are also
related to the profile of the gas disks. With higher gas density,
the speed of type I migration is higher. Planet pair is tend to be
captured into 3:2 MMR rather than 2:1 MMR \citep{Mustill2011,
Ogihara2013}. With flatter gas disk, planets are more easier to be
in 3:2 MMR \citep{Wang2017}.

Moreover, we propose a likely scenario for configuration formation
of the Kepler-68 system. As the eccentricity of $P_c$ and the mass
of $P_d$ are not well determined
\citep{Gilliland2013,VanEylen2015,Mills2019}, we infer that $P_c$
cannot be stirred up to a moderate value unless there is a massive
giant companion in the system. Future observations should be
addressed to further decode the origin of Kepler-68. The innermost
planet in this system is very close to the central star, the tidal
effect arising from the central star is a main reason leading to the
deviation from exact 5:3 MMR \citep{Lee2013}. Another possible
scenario is the eccentricity damping effect induced by the depletion
of the gas disk (Wang et al. in preparation).

Last but not least, our scenario can also be applied to formation of
those systems that closely resemble Kepler-68, e.g., Kepler-65,
Kepler-154 and Kepler-167 \citep{Kipping2016, Berger2018,
Mills2019}. For instance, Kepler-65 harbors three inner terrestrial
planets and a distant giant planet with an orbital period 258.8 day
and an eccentricity 0.28, where two planet pairs among them are near
2.7 and 1.4, respectively \citep{Chaplin2013}. According to our
formation scenario, we may infer that the eccentricities of three
terrestrial planets can be stirred up to about 0.1 by the giant
companion. During the subsequent migration, the outer planet pair
could be trapped into 7:5 MMR, whereas the inner planet pair may be
captured into 5:2 MMR as the outermost giant planet perturbs.
Furthermore, as the innermost planet orbits close to the central
star, its eccentricity may gradually decline by tidal effect from
the host star over secular timescale as well as the decrease of
semi-major axis, thereby producing the final period ratio between
the inner pair above 2.5. In conclusion, our model can throw light
on the formation of the planetary systems that harbor several
terrestrial planets near MMRs inside and one giant planet outside as
observed by Kepler mission.

\section*{Acknowledgements}
We thank the anonymous referee for constructive comments
and suggestions to improve the manuscript. This work is
financially supported by the B-type Strategic Priority Program of
the Chinese Academy of Sciences (Grant No. XDB41000000), the
National Natural Science Foundation of China (Grant Nos. 11773081,
11573073,11633009), CAS Interdisciplinary Innovation Team,
Foundation of Minor Planets of the Purple Mountain Observatory and
Youth Innovation Promotion Association.

\section*{Data availability}
The data underlying this article will be shared on reasonable request to the corresponding author.





\bsp    
\label{lastpage}
\end{document}